\documentclass[%
 aip,
 amsmath,amssymb,
 reprint,%
]{revtex4-1}
\usepackage[utf8]{inputenc}
\usepackage{amssymb,stackengine,scalerel}
\usepackage{mathtools}
\usepackage{hyperref}
\usepackage{helvet}
\usepackage{graphicx}
\usepackage{dcolumn}
\usepackage{bm}

\usepackage[utf8]{inputenc}
\usepackage[T1]{fontenc}
\usepackage{mathptmx}
\usepackage{etoolbox}
\usepackage{subfigure}
\usepackage{amsmath}
\usepackage{xcolor}
\newcommand{\rr}{\textcolor{black}}
\newcommand{\bbb}{\textcolor{black}}
\newcommand{\rrr}{\textcolor{black}}

\makeatletter
\def\@email#1#2{%
 \endgroup
 \patchcmd{\titleblock@produce}
  {\frontmatter@RRAPformat}
  {\frontmatter@RRAPformat{\produce@RRAP{*#1\href{mailto:#2}{#2}}}\frontmatter@RRAPformat}
  {}{}
}%

\makeatother
\begin{document}

\preprint{AIP/123-QED}

\title[]{Molecular Processes as Quantum Information Resources}
\author{Saikat Sur}
 \affiliation{${}^1$Department of Chemical \& Biological Physics,\\ Weizmann Institute of Science, Rehovot 7610001, Israel}
\author{Pritam Chattopadhyay}%
\affiliation{${}^1$Department of Chemical \& Biological Physics,\\ Weizmann Institute of Science, Rehovot 7610001, Israel}%

\author{Gershon Kurizki$^*$}
 \email{gershon.kurizki@weizmann.ac.il}
\affiliation{${}^1$Department of Chemical \& Biological Physics,\\ Weizmann Institute of Science, Rehovot 7610001, Israel}%

\date{\today}

\begin{abstract}
In this contribution to Abraham Nitzan's Festschrift, we present a perspective of theoretical research over the years that has pointed to the potential of molecular processes to act as quantum information resources. Under appropriate control, homonuclear dimer (diatom) dissociation (half-collision) and the inverse process of atom-pair collisions are shown to reveal translational (EPR-like) entanglement that enables molecular wavepacket teleportation. When such processes involve electronic-state excitation of the diatom, the fluorescence following dissociation can serve as an entanglement witness that unravels the molecular-state characteristics and evolution. Such entangling processes can also exhibit anomalous quantum thermodynamic features, particularly temperature enhancement of a cavity field that interacts with dissociated entangled diatoms.
\end{abstract}

\maketitle

\section{Introduction}

Einstein, Podolsky, and Rosen (EPR) introduced in 1935 a state in which two particles had a well-deﬁned sum of momenta and of positions difference~\cite{EPR}. This state was meant to illustrate what they viewed as the incompleteness of quantum theory. Shortly thereafter Schr{\"{o}}dinger~\cite{schrodinger1935discussion,schrodinger1935present} pointed out the ``paradoxical” features of states that describe two-particle entanglement of which the EPR state is an example. Even earlier, Von Neumann had elucidated~\cite{von2013mathematische} the informational aspects of entangled, namely, inseparable, two-particle states.

\rr{As an alternative to EPR states of translational continuous variables, Aharonov and Bohm~\cite{PhysRev.108.1070} put forth the entanglement of discrete spin-$\frac{1}{2}$ observables for atom pairs in Stern-Gerlach setups. Such spin-entangled states were subsequently used by Bell to formulate his inequality that can test the validity of quantum-mechanical non-locality~\cite{bell2004speakable}. }

Entanglement has since become the cornerstone of quantum physics, particularly 
 of quantum information science~\cite{PhysRevLett.92.127903,von2013mathematische,chan2004observable,eisert2002quantification,PhysRevA.66.042113,PhysRevA.61.032305,PhysRevLett.85.2657,nielsen2010quantum,steane1998quantum}. Yet, the original EPR scenario of continuous-variable entanglement has mainly been studied for quantum-correlated electromagnetic fields~\cite{chan2004observable,PhysRevLett.92.127903,PhysRevLett.80.869,PhysRevLett.60.2731,furusawa1998unconditional} or photon-pairs~\cite{PhysRevLett.92.210403} rather than for translationally-entangled free (unbound) \textit{massive particles}, whose EPR correlations have not been experimentally investigated to date. \rr{Likewise, the discrete-variable  Bell’s inequality was tested by converting  spin-$\frac{1}{2}$  entanglement to correlations between photons emitted by an atom~\cite{PhysRevLett.36.1223,PhysRevLett.49.1804}, and photon–polarization variables have since been used in most schemes of discrete-variable entanglement~\cite{scully1997quantum,meystre2007elements}.}
 
In contrast to \rr{the prevailing photon-based} EPR studies~\cite{chan2004observable,PhysRevLett.92.127903,PhysRevLett.80.869,PhysRevLett.92.210403,PhysRevLett.60.2731}, Opatrný and Kurizki~\cite{prl_86_3180_2001} proposed the creation of \textit{translationally-entangled pairs of free particles (unbound atom pairs)} approximating the EPR state by dissociation of cold diatomic molecules. \rr{Analogous processes were theoretically considered by Kurizki and Ben-Reuven~\cite{PhysRevA.32.2560,PhysRevA.36.90,ben1987stereospecificity} and experimentally demonstrated by Grangier, Aspect, and Vigu\'e~\cite{PhysRevLett.54.418} to yield \textit{discrete-variable entanglement} in two-atom states formed by diatomic molecular dissociation.  Here we present a broader perspective of molecular processes as natural resources of both continuous- and discrete-variable entanglement, whose relevance to quantum information has been largely overlooked by experimentalists. Their potential quantum-technological applications and challenges are the focus of this perspective.}

\rr{A major trend of quantum technologies has been the use of entanglement as a resource \textit{for quantum teleportation}: the fundamental prescription of how to uniquely map quantum states of one system (A) onto those of another (B), by measuring A and an auxiliary system in a joint entangled-state basis and then manipulating the corresponding observables in B according to the results of this measurement~\cite{PhysRevLett.80.869,furusawa1998unconditional,PhysRevLett.70.1895,PhysRevA.49.1473}.   The concept of quantum teleportation by Bennett et al. was inspired by the \textit{Startrek} replica ``Beam me up, Scottie”. Although quantum teleportation cannot be performed instantaneously, without prior sharing of the measurement results between the sender and the receiver at the speed of light, it offers, in principle,  the spectacular possibility of transferring an unknown quantum state of an object between distant nodes,  which would allow the recreation of this object arbitrarily far from the sender.}

\rr{Yet,  how can we teleport  (``beam up") the quantum states of material objects,  if the entangled and teleported states at our disposal are merely photonic?  Schemes enabling the entanglement of distant material systems as a step towards their teleportation have been devised for trapped cold-atom ensembles~\cite{PhysRevA.62.033809,julsgaard2001experimental},  and for nanomechanical systems~\cite{bouwmeester1997experimental,lamas2002experimental}.}


 The Opatrný-Kurizki proposal~\cite{prl_86_3180_2001} had offered a glimpse into the ability of \rr{molecular dimer dissociation to provide two-atom  EPR entanglement on demand and its use for matter-wave quantum teleportation. This teleportation scheme was later extended to translational EPR-like entanglement of atom-pairs that undergo molecular collisions~\cite{PhysRevLett.94.160503}. Similarly, the Kurizki-Ben Reuven theory~\cite{PhysRevA.32.2560,PhysRevA.36.90,ben1987stereospecificity} has laid the basis for preparing \textit{entangled  Bell states} of two-atom discrete variables via molecular dissociation and collisions. Their use for teleportation is yet to be explored.} 

\rr{The following questions, which are key to the validity of these molecular processes as \textit{translational entanglement} resources, are discussed} in Sec.~\ref{section_2} to \ref{section_4}:

(1) What are the adequate criteria or measures for translational entanglement in a collision or half-collision (dissociation)?

(2) How to prevent (or mitigate) translational entanglement loss caused by wavepacket broadening in the course of collisions or dissociation?

(3) How to measure coordinate and momentum correlations in such processes so as to verify their EPR entanglement?


\rr{\textit{Discrete-variable (internal-state) entanglement} between the products of molecular dissociation or collision is discussed in Sec.~\ref{section_5}, based on cooperative two-atom }
 features of fluorescence at dissociation-product (fragment) separations well beyond those of molecular forces~\cite{PhysRevA.32.2560,PhysRevA.36.90,ben1987stereospecificity,PhysRevA.38.6433,PhysRevLett.54.418}. As we point out, the time-resolved fluorescence at such separations is a \emph{witness of quantum entanglement} in the dissociated molecule through correlations between photon-emissions of the fragments, since these correlations are determined by the entangled state of the parent molecule (Sec.~\ref{section_5}). \rr{Therefore, these time-resolved fluorescence patterns can be an alternative tool for identifying/diagnosing the parent-molecule states and their dynamics. }

The translationally-entangling processes discussed in Sec.~\ref{section_2}-\ref{section_4} are shown in Sec.~\ref{section_6.1} to enable \textit{molecular wavepacket teleportation} by performing two-particle position and momentum measurements~\cite{PhysRevLett.80.869,prl_86_3180_2001,fisch2005translational,fisch2006free} on one of the products of a collision or half-collision. \rr{Discrete-variable teleportation may be based  on Bell measurements of the fragments that are  entangled as in Sec.~\ref{section_5.1}.} 

\rr{ A much less familiar class of processes surveyed in Sec.\ref{section_5.2} are those that create \textit{translationally-internally entanglement (TIE)} whereby discrete internal variables of a system are correlated to its continuous (translational) variables, as proposed for single atoms in interferometers~\cite{kolavr2007path} and demonstrated for atom pairs in a cold gas~\cite{gross2011atomic}.} In Sec.~\ref{section_6.2} we discuss a scheme based on stimulated Raman adiabatic passage (STIRAP)~\cite{RevModPhys.70.1003} for transferring \rr{TIE} states from molecules in a cavity to two-mode photonic fields~\cite{PhysRevA.67.012318}, which can subsequently recreate, upon reversing the process, the same entangled state in a molecule located in a distant cavity (Sec.~\ref{section_6.2}).\\

Finally, we point (Sec.~\ref{section_7}) to an anomalous quantum thermodynamic feature of dissociating dimers namely, their \textit{entanglement-dependent temperature}~\cite{dag2019temperature} and compare it to the temperature dependence on \textit{single-atom coherence} predicted in a pioneering paper by Scully et al.~\cite{scully2003extracting}. 

The Discussion (Sec.~\ref{section_8}) highlights the main points of the above processes and identifies \rr{conceptual as well as } technical hurdles, \rr{particularly those related to measurements that are indispensable in quantum information schemes~\cite{furusawa1998unconditional,PhysRevLett.92.210403}.  These issues need to be addressed for the realization of the surveyed protocols and their extension to more complex systems.} 

\rr{A note on the terminology that underlies most of the content: Readers unfamiliar with molecular dynamics and scattering theory are advised to consult not only the cited articles but also textbooks, such as  Davidov’s \textit{Quantum  Mechanics} (Ch. XIV-XV )~\cite{davydov2013quantum} or refs.~\cite{baer2012molecular,child2014molecular}. The figures in this perspective were inspired by (but not copied from) those in the corresponding cited articles.}


  \begin{figure*}
 \begin{center}
 \subfigure[]{%
            \includegraphics[width=0.365\textwidth]{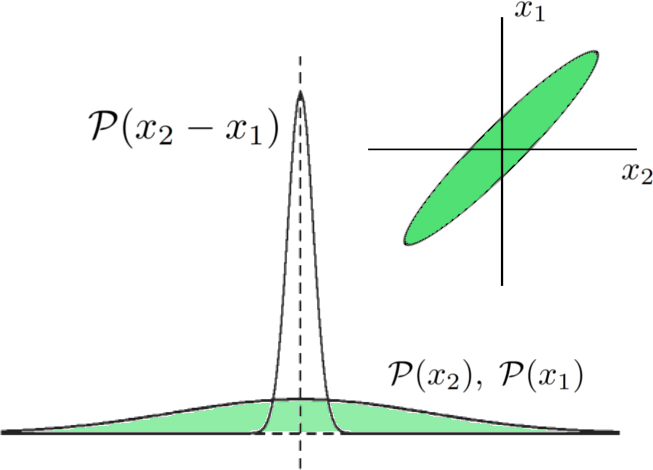}
        }
                 \subfigure[]{%
            \includegraphics[width=0.36\textwidth]{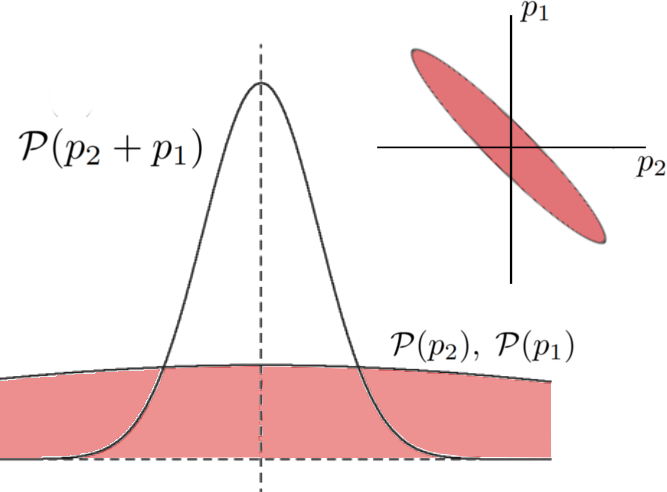}
        }\\%
 \end{center}
 \caption{Probability distribution and phase-space contour of (a) distance $\mathcal{P}(x_2 -x_1)$ and (b) the sum of momenta $\mathcal{P}(p_2 + p_1)$ of an EPR pair, and their single-particle counterparts $\mathcal{P}(x_{(1)})$ and $\mathcal{P}(p_{(1)}).$}
 \label{fig_1} 
 \end{figure*}

  \begin{figure*}
 \begin{center}
 \includegraphics[width=0.46\textwidth]{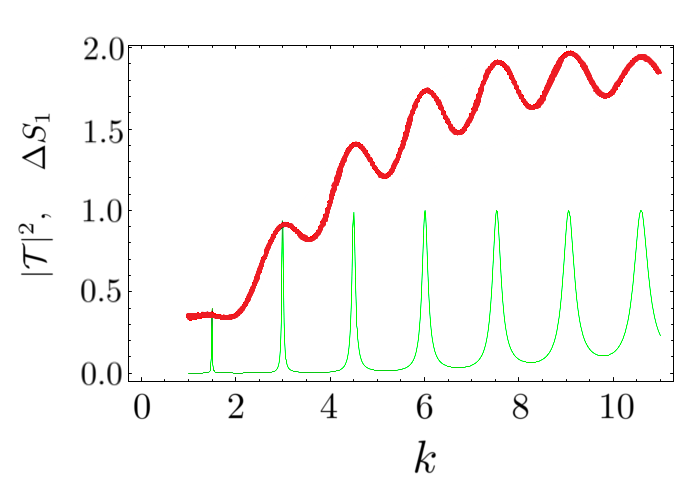}

 \end{center}
 \caption{Entanglement quantified by change in single-particle VN entropy, $\Delta S_1$, in 1d collisions via a double $\delta$-function potential localized at $x = \pm a$ as a function of the relative momentum $k$ for wavepacket momentum width $\Delta k \gg \sigma$ (thick red curve).  Thin green peaks indicate the resonances of the transmission coefficient,
the peaks corresponding to $|\mathcal{T}|^2 = 1$.
}
 \label{fig_2} 
 \end{figure*}

   \begin{figure*}
 \begin{center}
          \subfigure{%
            \includegraphics[width=0.56\textwidth]{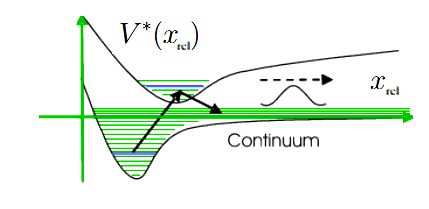}
        }\\%
  \end{center}
 \caption{EPR entanglement of ground state (S+S) formed by Raman-induced dissociation via the electronically-excited potential $V^*_{(x_{\text{rel}})}$ followed by post-dissociative spreading of $x_{\text{rel}}$ distribution in the continuum. } 
 \label{fig_3} 
 \end{figure*}

   \begin{figure*}
 \begin{center}
  \includegraphics[width=0.52\textwidth]{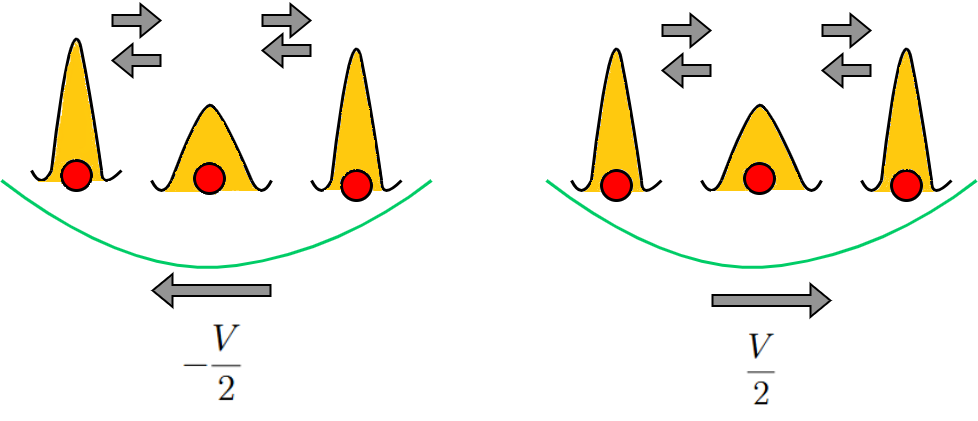}
 \end{center}
 \caption{Confinement of receding dissociating fragments in time-dependent double-well trap. Double arrows indicate quarter-period oscillation of $x_1$, $x_2$ wavepackets for fragments confined in receding traps.}
 \label{fig_4} 
 \end{figure*}

\section{ EPR States and Translational Entanglement Measures}
\label{section_2}
The two-particle EPR states with ideal coordinate and momentum entanglement in one dimension (1d) are defined as

\begin{eqnarray}
 \left\langle x_{1}, x_{2} \mid E P R_{\mp}\right\rangle & =\delta\left(x_{1} \mp x_{2}\right) \nonumber\\
\left\langle p_{1}, p_{2} \mid E P R_{\mp}\right\rangle & =\delta\left(p_{1} \pm p_{2}\right),
\label{eq: state_1}  
\end{eqnarray}
where $x_{1(2)}$ and $p_{1(2)}$ denote the  $1(2)$ particle positions and momenta. These EPR states  are \emph{unphysical}, being \emph{unnormalizable} and having infinite energy spread, but they are the limits of the physical gaussian state $\vert \psi \rangle$, which has the following coordinate representation

\begin{eqnarray}\label{eqnn2}
\left\langle x_{1}, x_{2} \mid \psi\right\rangle=N e^{-\left(\frac{x_{\mathrm{cm}}}{2 \Delta x_{\mathrm{cm}}}\right)^{2}} e^{-\left(\frac{x_{\mathrm{rel}}}{2 \Delta x_{\mathrm{rel}}}\right)^{2}},
\label{eq: state_2}
\end{eqnarray}
where $N$ is the normalization, $x_{\text {rel }} \equiv x_{1}-x_{2}$ and $x_{\mathrm{cm}} \equiv \frac{x_{1}+x_{2}}{2}$ being respectively the relative and center-of-mass positions. As $\Delta x_{\text {rel}} \rightarrow 0, \Delta x_{\mathrm{cm}} \rightarrow \infty$ Eq.~\eqref{eqnn2} reduces to the ideal $\mathrm{EPR}_{-}$ state and the limit $\Delta x_{\text {rel }} \rightarrow \infty, \Delta x_{\mathrm{cm}} \rightarrow 0$ to its $\mathrm{EPR}_{+}$ counterpart (Fig.~\ref{fig_1}). In such states the positions or momenta of the two particles along the $x$ axis are completely uncertain, but perfectly correlated or anti-correlated: $x_1  = x_2, p_1 = -p_2$ or vice versa.

The proximity of the gaussian state (\ref{eq: state_2}) to the ideal EPR states (\ref{eq: state_1}) depends on the extent to which the variances of the correlated two-particle variables are below the single-particle Heisenberg uncertainty limit: e.g., for $x_{rel}$ and the center-of-mass momentum $\text{p}_{cm}$ 

\begin{eqnarray}
   \Delta x_{rel} \Delta p_{cm} \ll \hbar. 
\end{eqnarray}
Since these variables commute $[ x_{rel}, p_{cm}]=0$, this inequality is consistent with quantum mechanics but not with the EPR notion of ``elements of reality"~\cite{EPR}. 

The proximity to ideal EPR states can be quantified by several alternative criteria:\\
(a) Opatrný and Kurizki proposed~\cite{prl_86_3180_2001} the two-particle matter-wave \textit{``squeezing" parameter $s$}, defined as

\begin{eqnarray}\label{eqn3}
s(t)=\frac{\hbar}{2 \Delta x_{1}^{(c)}(t) \Delta p_{1}^{(c)}},
\label{eq_3}
\end{eqnarray}
$\Delta x_{1}^{(c)}$ being  the variance of the conditional probability distribution

\begin{eqnarray}
\mathcal{P}\left(x_{1} \mid x_{2}=a\right)=\frac{\left|\psi\left(x_{1}, x_{2}=a\right)\right|^{2}}{P\left(x_{2}=a\right)}
\end{eqnarray}

where $P\left(x_{2}=a\right)=\int d x_{1}\left|\psi\left(x_{1}, x_{2}=a\right)\right|^{2}$, $a$ being a chosen position. Analogously, $\Delta p_{1}^{(c)}$ is the conditional momentum uncertainty.

We may now define $s(t)$ for the wavepackets in Eq.~(\ref{eqn3}): Upon approaching the $\mathrm{EPR}_{+}$ limit, we have $\Delta x_{1}^{(c)}=\min \left(2 \Delta x_{\mathrm{cm}}, \Delta x_{\mathrm{rel}}\right)$, whereas near the EPR\_ limit, $\Delta p_{1}^{(c)}=$ $\min \left(2 \Delta p_{\mathrm{cm}}, \Delta p_{\mathrm{rel}}\right)$. From the above discussion, it follows that the parameter $s$ is a legitimate measure of the EPR correlation for both EPR states. 
We may define the EPR regime as $s>1$.  For a free-evolving $|\psi\rangle$ in Eq.~(\ref{eq: state_2}), the uncertainties in the conditional variance in (\ref{eqn3}) grow as $\Delta x^{(c)}_{j}(t)^{2}=\Delta x^{(c)}_{j}(0)^{2}+\Delta {p^{(c)}}^2_{j} t^{2} / m^{2}$, so that $s(t)$ diminishes with time.\\

The EPR regime,  $s > 1$ is currently achievable for pairs of photons created via parametric downconversion, upon combining near-field measurements of $\Delta x_{\text {rel }}$ with far-field measurements of $\Delta p_{\mathrm{cm}}$~\cite{PhysRevLett.92.210403}. By contrast, it is unclear how to prepare and measure  EPR correlations between pairs of \textit{matter waves}, formed by dissociation or a collision. These issues are discussed below.

(b) \textit{The Schmidt number $K$}  that has been adopted as an entanglement measure of a gaussian state of the form (\ref{eq: state_2}) satisfies~\cite{PhysRevLett.78.3221,parkins1999quantum}

\begin{equation}\label{eqnn6}
K=\frac{\Delta p_{1}}{\Delta p_{1}^{(c)}} 
\end{equation}

(c) One may quantify EPR entanglement by the \textit{Von-Neumann (VN) entropy} $S$, associated with the reduced density matrix of a single particle with discrete eigenvalues $\epsilon_n$~\cite{breuer2002theory} 

\begin{subequations}\label{eqnnnnnn5}
\begin{eqnarray} \label{eqnnnnnn7a}
S \equiv-\sum_{n=0}^{\infty} \epsilon_{n} \ln \epsilon_{n}
\end{eqnarray}

The evaluation of the VN entropy $S$ for unbound-particle collisions is nontrivial. For the gaussian state  (\ref{eq: state_2}), $S$ reduces
in the limit $\Delta x_{\mathrm{cm}} \gg \Delta x_{\mathrm{rel}}$ to  

\begin{eqnarray}
 S \simeq \log [2 \Delta x_{\mathrm{cm}}(t) / \Delta x_{\mathrm{rel}}(t)]=\log s(t).   
\end{eqnarray}
\end{subequations}
Equations~\eqref{eqnn6}, \eqref{eqnnnnnn5} are measures of translational entanglement of an arbitrary \emph{(not necessarily gaussian) bipartite} wavepacket. These measures are required for the description of collisions between unbound particles (Sec.~\ref{section_3}).

\section{Translational EPR Entanglement via Collisions}
\label{section_3}

It is generally complicated to quantify translational entanglement from the full scattering analysis of collisions~\cite{baer2012molecular,child2014molecular}. Tal and Kurizki~\cite{PhysRevLett.94.160503} substantially simplified the problem by assuming that each initial momentum state $\left|\mathbf{k}_{i}\right\rangle$  scatters onto a \emph{discrete}, orthonormal set of final momentum states with momenta $\vert\mathbf{k}_{f}\rangle$. Such discretization is adequate for momentum wavepackets having small widths near these discrete values, as detailed below. 

In this approximation, the initial two-particle state is taken to be an entangled superposition in the c.m. frame

\begin{eqnarray}
\left|\psi_{i}\right\rangle=\sum_{i} b_{i}\left|\mathbf{k}_{i}\right\rangle \otimes\left|-\mathbf{k}_{i}\right\rangle. 
\end{eqnarray}


In the post-collision density matrix obtained by tracing out particle 2, the single-particle (reduced) post-collision density operator is $\left.\left(\rho_{1}\right)_{f}=\sum_{\mathbf{k}_{f}}\left|\sum_{i}\left\langle\mathbf{k}_{f}\right| \mathbf{S}\right| \mathbf{k}_{i}\right\rangle\left.\right|^{2}\left|\mathbf{k}_{f}\right\rangle\left\langle\mathbf{k}_{f}\right|$, where $\mathbf{S}$ is the scattering $S-$ matrix. 

The initial single-particle entropy is $\left(S_{1}\right)_{i}=-\sum_{i}\left|b_{i}\right|^{2} \log _{2}\left|b_{i}\right|^{2}$. Then, to lowest order in the momentum widths of the pre-collision $\mathbf{k}_i$ components, the VN entropy change $\triangle S_{1}^{(0)}$ of each particle following the collision is:

\begin{eqnarray}\label{eqn8}
\Delta \mathbf{S}_{1}^{(0)}&=&-\sum_{\mathbf{k}_{f}}\left|\sum_{i} b_{i} \mathbf{S}_{\mathbf{k}_{\mathbf{f}}, \mathbf{k}_{\mathbf{i}}}\right|^{2} \log _{2}\left(\left|\sum_{i} b_{i} \mathbf{S}_{\mathbf{k}_{\mathbf{f}}, \mathbf{k}_{\mathbf{i}}}\right|^{2}\right)\nonumber\\&&+\sum_{i}\left|b_{i}\right|^{2} \log _{2}\left|b_{i}\right|^{2},
\end{eqnarray}
where $\mathbf{S}_{\mathbf{k}, \mathbf{k}^{\prime}} \equiv\langle\mathbf{k}| \mathbf{S}\left|\mathbf{k}^{\prime}\right\rangle$. 

The second term on the r.h.s of Eq.~\eqref{eqn8} yields the classical Boltzmann law for entropy change in multi-channel collisions. By contrast, the first term accounts for quantum interferences of different scattering channels and their entanglement if two or more $b_{i} \neq 0$, rendering $\triangle S_{1}$ \textit{non-classical}.

This collisional entanglement is maximized near scattering resonances. In 1d collisions, the  $S$-matrix effect on single-particle momentum states is given by
\begin{equation}
\mathbf{S}|k\rangle=\mathcal{T}(k)|k\rangle+ \mathcal{R}(k)|-k\rangle, 
\end{equation}
the first and second terms denoting the transmission and reflection caused by the interaction potential $V\left(x_{\text {rel }}\right)$, respectively.

The collisional entanglement near resonances can be evaluated by expanding the transmitted and the reflected portions of the single-particle post-collision density matrix, $\rho_{1}^{T}$ and $\rho_{1}^{R}$ to second order in the momentum spread $\Delta k^{2} \equiv\left\langle k-k_{0}\right\rangle^{2}$. Then the VN entropy change (cf.~\eqref{eqnnnnnn7a}) can shown to satisfy~\cite{PhysRevLett.94.160503}

\begin{subequations}

\begin{equation}
\Delta S_{1}^{(2)}  =-\epsilon_{T}^{(2)} \log _{2} \epsilon_{T}^{(2)}-\epsilon_{R}^{(2)} \log _{2} \epsilon_{R}^{(2)},
\end{equation}
where the transmitted- and reflected-wave scattering eigenvalues are given, to second order in $\Delta k$, by

\begin{equation}
\epsilon_{T}^{(2)}  =\left|\mathcal{T}\left(k_{0}\right)\right|^{2}+\frac{\Delta k^{2}}{4} \frac{d^{2}}{d k^{2}}|\mathcal{T}(k)|_{k=k_{0}}^{2}    
\end{equation}

\begin{equation}
\epsilon_{R}^{(2)}=\left|\mathcal{R}\left(k_{0}\right)\right|^{2}-\frac{\Delta k^{2}}{4} \frac{d^{2}}{d k^{2}}|\mathcal{T}(k)|_{k=k_{0}}^{2}
\end{equation}
    
\end{subequations}

Let the wavepacket be initially confined, in $k$-space, to a 1 d region of width $\Delta k \sim \sigma$. After the collision, the wavepackets are modulated in $k$ by $\mathcal{T}$ and $\mathcal{R}$, whose scale of change is given by $\Gamma$. The post-collision wavepackets are then confined in $x-$ space to a region $\Delta x_{j} \sim \frac{1}{\Gamma}$. This yields the VN entropy estimate
\begin{equation}
\Delta\left(S_{j}\right)_{\max } \sim \log _{2}\left(\frac{\sigma}{\Gamma}\right)(j=1,2).
\end{equation}

More detailed analysis, supported by numerics~\cite{PhysRevLett.94.160503,fisch2005translational}, shows that initially narrow momentum wavepackets yield double peaks of the entanglement on both sides of a resonance. The range $\sigma \ll \Gamma$ gives rise to a dip in $\Delta S_1$ at the transmission resonance $\vert \mathcal{T}(k) \vert^2 = 1$. By contrast, $\Delta S_1$ is peaked at this resonance for $\sigma \gtrsim \Gamma$. This entropy change further grows for $\Delta k \gg \sigma \gg \Gamma$, where the \emph{wavepacket approximates the EPR state} (Fig.~\ref{fig_2}). This simple model of collisional entanglement gives qualitative insights into the ability of elastic collisions to create EPR correlations between the two atoms.

\section{Means of Preserving and Measuring EPR Correlations}
\label{section_4}
\subsection{Molecular Raman Dissociation}
\label{section_4.1}

For quantifying the EPR correlations due to a collision or half-collision one should be able to estimate the center-of-mass (cm) and relative (rel) two-particle wavepackets $\left|\psi_{\mathrm{cm}}\right\rangle$ and $\left|\phi_{\text {rel }}\right\rangle$ respectively, in order to infer their position and momentum uncertainties along a particular axis, at asymptotically large separations between the particles $r_{\text {rel }}$.

At long times $t$ after the onset of a collisional process, an initial bipartite wavepacket evolves, within the stationary phase approximation, into
$\vert \phi_{\text{rel}} (t=0)\rangle$  of two particles with mass $m$ and relative energy $E_{\text{rel}}$

\begin{eqnarray}
&& \left\langle\mathbf{r}_{\mathrm{rel}} \mid \phi_{\mathrm{rel}}(t)\right\rangle \rightarrow \sum_{l,m} \int d E e^{-i \frac{\hbar \hbar^{2} t}{2 m}} c_{E l m}(t=0) \nonumber\\
&& \quad \times \frac{1}{2 i k r}\left(e^{i\left(k r-\frac{\pi l}{2}\right)}-e^{-i\left(k r-\frac{\pi l}{2}\right)}\right) Y_{l m}(\theta, \phi).
\end{eqnarray}
where $k=\sqrt{2 m E_{\text {rel }} / \hbar}$ and $c_{E l m}$ are the initial amplitudes of   the eigenstates  $\vert Elm\rangle$ of the full (scattering) Hamiltonian.  The asymptotic  $\left|\phi_{\text {rel }}\right\rangle$ formed by a \emph{spherical interaction potential} is given by

\begin{eqnarray}
\left|\phi_{\mathrm{rel}}(t \rightarrow \infty)\right\rangle &\sim& \sum_{l, m} \int d E\left\langle Elm \mid \phi_{\mathrm{rel}}(0)\right\rangle \nonumber\\ &&\times e^{i \delta_{l}(E)} e^{\frac{-i E t}{\hbar}}\left|{E l m}\right\rangle_0.
\end{eqnarray}
Here, $\delta_{l}(E)$ are the phase shifts of the partial scattered waves, $\left|E l m\right\rangle_0$ and $\left|E l m\right\rangle$ being the respective eigenstates of the free and scattering Hamiltonians $H_{0}$ and $H$. 

In what follows we discuss the coordinate and momentum spread in molecular Raman dissociation~\cite{akulin2012intense}, which is advantageous for EPR entanglement. 
In Raman dissociation of a diatom, laser beams cause a transition from the initially bound ground state to an unbound state in the continuum, through an intermediate bound state on an excited electronic surface (Fig.~\ref{fig_3}). The scattered relative-motion wavepacket of the two receding atomic fragments is then approximately described by~\cite{fisch2005translational}  \\

\begin{eqnarray}\label{eqn15}
  \phi_{\mathrm{rel}}\left(\mathbf{r}_{\mathrm{rel}}, t\right) &\sim & \sum_{l=0,2} \sqrt{\frac{2 l+1}{4 \pi}} P_{l}(\cos \theta) \times \frac{  e^{i\left(k r_{\mathrm{rel}}-\frac{\pi l}{2}+\delta_{l}(E)-E t / \hbar\right).}}{2 i k r_{\mathrm{rel}} } \nonumber\\
&& \times \theta\left(v t-r_{\mathrm{rel}}\right) \sin \left(\frac{\Omega_{\mathrm{eff}}}{2}\left(t-\frac{r_{\mathrm{rel}}}{v}\right)\right),
\end{eqnarray}
where $\Omega_{eff}$ is the effective Rabi frequency of the Raman transition. 
This expression is a first-order perturbation-theory approximation for an expanding wavefunction with an edge at $r_{\text {rel }}=v t$.

The laser pulse that induces the Raman transition should minimize $\Delta E_{\mathrm{rel}}$, the energy width of the two-particle relative-motion wavepacket, and thereby minimize their relative momentum spread $\Delta p_{\text {rel }}$. The energy variance $\Delta E_{\text {rel }}$ scales with $ \Omega_{eff}$ and with the inverse time scale. Yet, $\Delta E_{\text {rel }}$ has a lower bound 
determined by the probability of radiative decay from the excited state which grows with $1/\Delta E_{\mathrm{rel}}$ Nevertheless, Raman dissociation induced by nearly-monochromatic laser beams, can yield an $\left|E P R_{+}\right\rangle$-like state, characterized by $s=\Delta x_{\mathrm{rel}} /2 \Delta x_{\mathrm{cm}} \gg 1$, but corrupted by small radiative decay.

\subsection{Storing the dissociating 2-particle wavefunction by receding harmonic wells}
\label{section_4.2}

The dispersion and spread of the two-atom wavepacket weaken their position and momentum correlations caused by dissociation or collisions. To preserve the initial degree of entanglement one can confine the two atoms near $x_{1,2} = \pm x_0/2$, in a double-parabolic trap (tweezer). The two-particle potential term in the Hamiltonian is then

\begin{eqnarray}\label{eqn16}
    V(x_1,x_2) = \frac{1}{2} \Big[M \omega^2 x^2_{c.m.} +  \mu \omega^2 (x_{rel} - x_0)^2\Big]
\end{eqnarray}
Here $M = 2m$ is the combined mass and $\mu = m/2$ the reduced mass of the atom pair.

Both $\vert \psi_{c.m.} \rangle$ and $\vert \phi_{rel} \rangle$ behave as \emph{single-particle wavefunctions} in a parabolic potential. The separability of $x_{cm}$ and $x_{\text{rel}}$ motion in Eq.~\eqref{eqn16} implies that if we separate the potential wells at the relative dissociation velocity $v = \langle p_{rel} \rangle/\mu $, each atom is kept in another well while their coordinate and momentum variances periodically oscillate within the wells (Fig.~\ref{fig_4}).   Their initial two-atom ``squeezing” in Eq.~\eqref{eqn3} is then preserved provided the two wells are turned on and start receding immediately after dissociation until the atoms are measured.

\section{Fluorescence in Dimer Dissociation As Entanglement Witness}
\label{section_5}
\subsection{Fluorescence from spinless atomic fragments in dimer dissociation}
\label{section_5.1}

 \begin{figure*}
 \begin{center}

         \subfigure[]{%
            \includegraphics[width=0.56\textwidth]{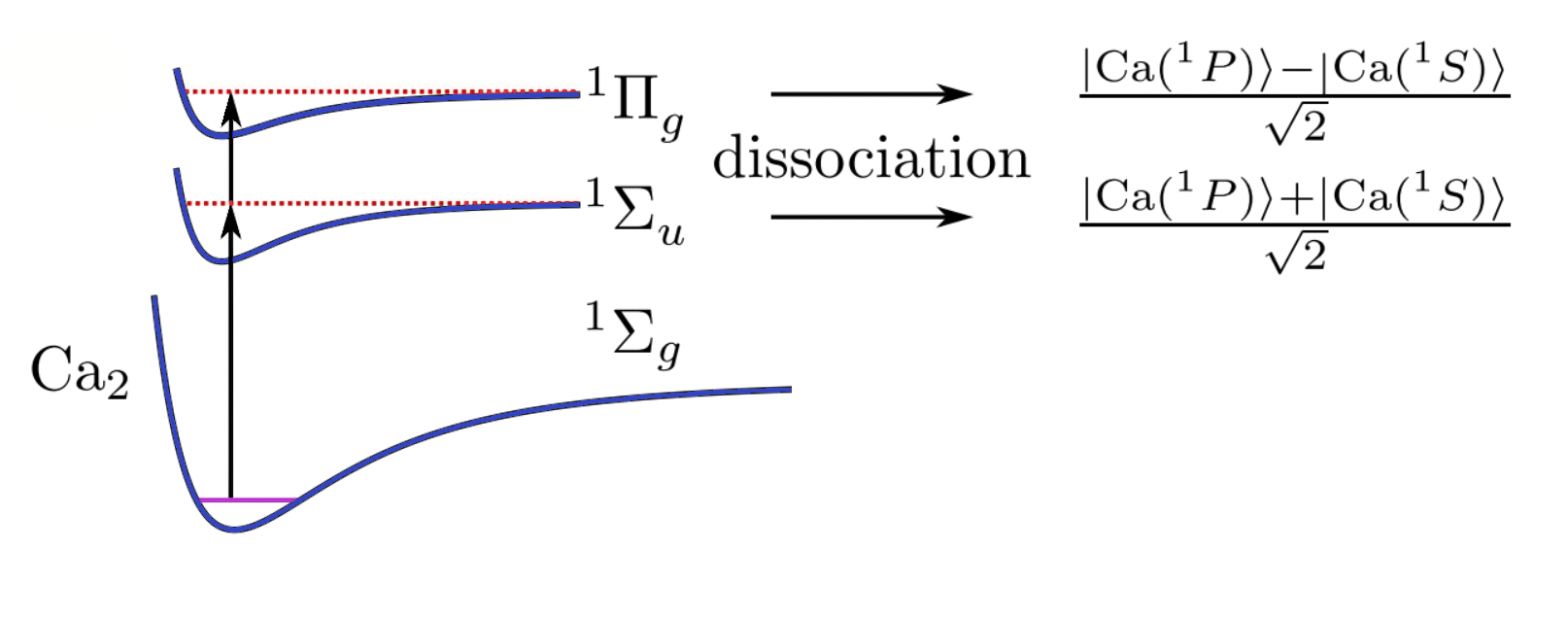}
        }\\%
                 \subfigure[]{%
            \includegraphics[width=0.7\textwidth]{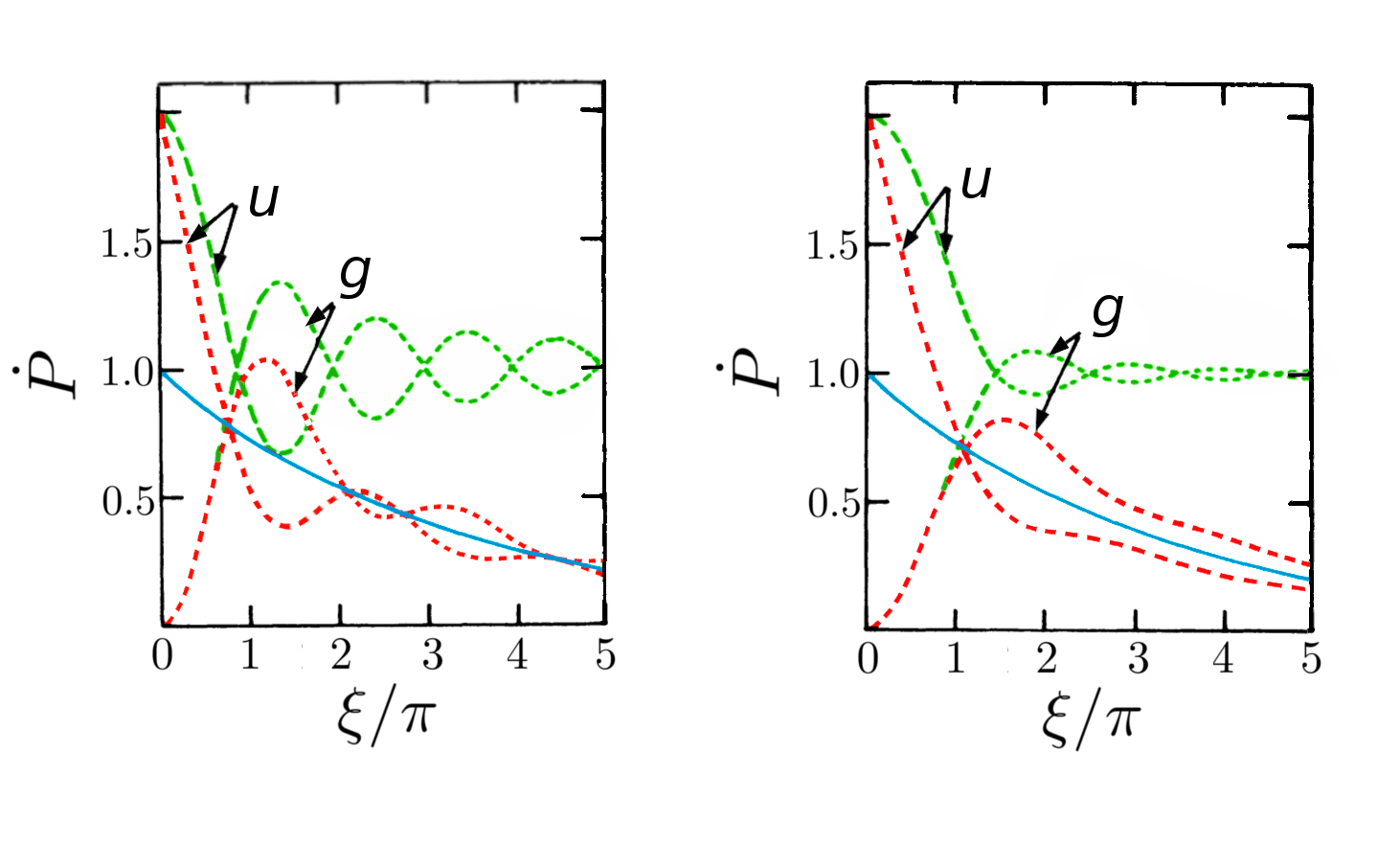}
        }\\%
 \end{center}
 \caption{(a) Single- or two-photon excitation taking $\text{Ca}_2$ above the dissociation threshold yields respectively, a symmetric (superradiant) $(^1\Sigma_u)$ or antisymmetric (subradiant) $(^1\Pi_g)$ Dicke state. (b) Left: Two-atom cooperative emission rate ($\dot{P}$) (in units of single-atom decay rate $\gamma$) as a function of  $\xi = k \dot{R}t$,  for $^1\Pi^*_u$ or a $^1\Pi^*_g$ dimer state dissociating into two singly-excited atoms ($\vert \Delta \Lambda\vert = 1$ transition).  Thin (blue) solid line --- single-atom emission rate $\gamma \exp(-\gamma t)$. Right: same, for a dimer prepared in a $^1\Sigma^*_u$ or a $^1\Sigma^*_g$ state ($\vert \Delta \Lambda\vert = 0$ transition).  The two-atom cooperative emission for $|\Delta \Lambda|=1$ and  $\Delta \Lambda=0$ transitions, respectively, is plotted for $\gamma / \dot{\xi}=0.1$ (red) and 0 (green). The two plots exhibit the ringing behavior for the Dicke $|1,0\rangle$ (u) and $|0,0\rangle$ (g) states. }
 \label{fig_5} 
 \end{figure*}

   \begin{figure*}
 \begin{center}
  \includegraphics[width=0.6\textwidth]{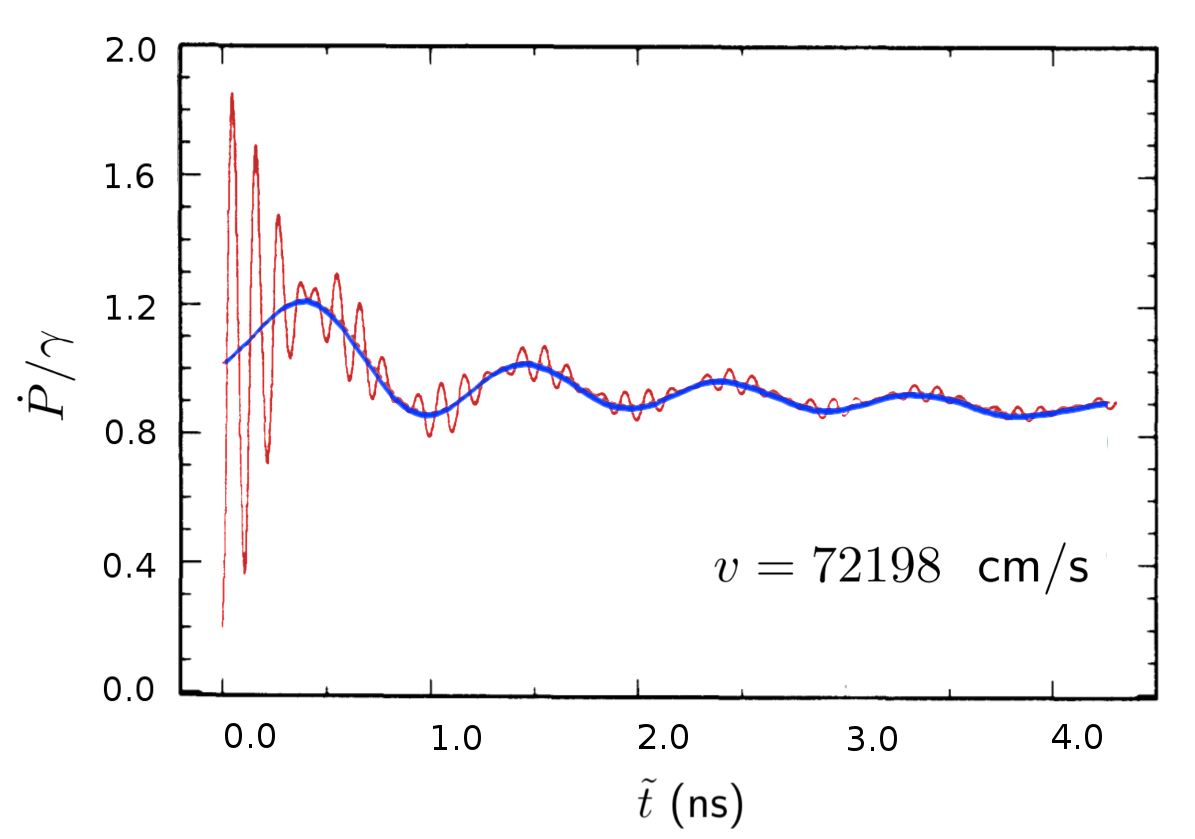}
 \end{center}
 \caption{Time-dependence of the fluorescence rate in units of single-atom rate (red curve) from a pair of atoms formed by Li$_2$ diatoms excited to
the continuum of the $1 {}^1\Sigma^+_u$ state by one photon: $\lambda_{ex} = 4291.89 $ {\AA}, where they recede at velocity $v$. Non-adiabatic dynamics imposes entanglement-induced fast fine-structure beats on the adiabatic-state ringing envelope that is analogous to Fig.~\ref{fig_5}(b).} 
 \label{fig_6} 
 \end{figure*}



In systems of two identical spinless atoms formed by dissociation of a molecule such as $\mathrm{Ca}_{2}$, the dissociation proceeds via a singly-excited electronic state, ${ }^{1} \Sigma_{u(g)}^{*}$ or ${ }^{1} \Pi_{u(g)}^{*}$. An ungerade (u) or gerade (g) molecular-symmetry state then yields at large internuclear separations a symmetric (antisymmetric) combination of states of atoms $a$ and $b$ $\left|{ }^{1} P_{m}\right\rangle_{a(b)}\left|{ }^{1} S\right\rangle_{b(a)}$. This scenario conforms to a model of two indistinguishable two-level dissociation fragments at large interatomic separations, where they undergo time-resolved ${ }^{1} P^{*} \rightarrow{ }^{1} S$ single-photon emission of fluorescence. 

The features of the fluorescence are determined, on the one hand, by the parity and total spin of the entangled electronically-excited state of the parent molecule, and,  on the other hand, by the electronic (spin and orbital) angular momentum states of the fragments. 
The time-resolved emission rate exhibits ringing in time,  reflecting
the interference between the emissions of the receding
fragments sharing the entangled state~\cite{PhysRevA.36.90,ben1987stereospecificity}. Such ringing has been experimentally demonstrated for the time-resolved emission from photo-dissociated
$\text{Ca}_2$ molecules~\cite{PhysRevLett.54.418}. 




The identical spinless two-level atomic fragments
$(a \text{~~or~~} b)$ with states $\vert e_{a(b)} \rangle, \vert g_{a(b)} \rangle$ can form the following Dicke-triplet states:

\begin{eqnarray}
   && \vert 1,1 \rangle = \vert e_ae_b, \rangle, \nonumber\\
   && \vert 1,0 \rangle = (1/\sqrt{2})(\vert e_ag_b \rangle + \vert g_ae_b \rangle),\nonumber\\
      && \vert 1,-1 \rangle =  \vert g_ag_b \rangle, \nonumber\\
\end{eqnarray}
or a Dicke singlet
\begin{eqnarray}
&& \vert 0,0 \rangle = (1/\sqrt{2})(\vert e_ag_b \rangle - \vert g_ae_b \rangle),\nonumber\\
\end{eqnarray}
where we have used the collective Dicke-state pseudospin notation $\vert s,s_z \rangle$~\cite{PhysRev.93.99}.

For $\vert e \rangle \rightarrow \vert g \rangle$ optical transitions in the atomic fragments, the following
correspondence exists between the entangled Dicke states and molecular inversion symmetry (parity) (Fig.~\ref{fig_5}(a))~\cite{PhysRevA.32.2560,PhysRevA.36.90,ben1987stereospecificity}: i) for even molecular spin (spin singlet):

\begin{subequations}
    \begin{eqnarray}
   && \vert 1,0 \rangle \leftrightarrow \text{ungerade~~}(u), \nonumber\\
   && \vert 0,0 \rangle \leftrightarrow \text{gerade~~}(g),\nonumber\\      \end{eqnarray}
  ii) whereas for odd molecular spin (spin triplet):
    \begin{eqnarray}
   && \vert 1,0 \rangle \leftrightarrow \text{gerade}, \nonumber\\
   && \vert 0,0 \rangle \leftrightarrow \text{ungerade}.\nonumber\\      \end{eqnarray}
\end{subequations}
If we detect the fluorescence soon after dissociation begins, we should find that the Dicke-triplet entangled state $\vert1,0\rangle$ yields ringing that corresponds to initially  \emph{constructive interference} of the single-atom emissions, while the Dicke-singlet entangled state $\vert0,0\rangle$ leads to their initially destructive interference (Fig.~\ref{fig_5}). 

In more detail, the total emission rate for such a singly-excited system is given by~\cite{PhysRevA.32.2560,PhysRevA.36.90,ben1987stereospecificity}
\begin{eqnarray}
\dot{P}(t) &=& [\gamma + \gamma(t)] \rho_{10,10}(t ) + [\gamma - \gamma(t)] \rho_{00,00}(t ) \nonumber\\
 &=& \dot{\rho}_{10,10}(t) + \dot{\rho}_{00,00}(t)
 \label{eq: 4}
\end{eqnarray}
The time-dependence here scales with $\xi(t) = kR(t)$ where $k$ is the emission wavevector and $R(t)$ is the separation which is usually given by $R(t) = vt$, $v$ being the dissociation velocity. Initially, i.e. at  short separations, we have $\gamma(t) \rightarrow \gamma$, so that the Dicke triplet \emph{superradiates} (at double
the emission rate of a single atom), while the
Dicke singlet is totally ``dark" (\emph{fully subradiant}). At later times, $\gamma(t)$ oscillates and so does the emission rate as shown in (Fig.~\ref{fig_5}).

The explicit solution depends  on whether one
starts from a $\Sigma^*$ or a $\Pi^*$ state of the dimer~\cite{ben1987stereospecificity}: 

a) For $\Sigma^* \rightarrow \Sigma (\Delta \Lambda = 0)$ transitions,
assuming the dissociation velocity $\Dot{R} = v$ to be constant, the rates in Eq.~(\ref{eq: 4}) are

\begin{eqnarray}
    &&\gamma \pm \frac{1}{t}\int^t_0 \gamma_{\Delta \Lambda = 0 }(t') dt' \nonumber\\&&= \frac{\gamma}{t \dot{\xi}}\big[ \xi \pm \frac{3}{2}[\text{Si}( \xi )- \xi^{-2} \sin \xi 
 + \xi^{-1} \cos \xi]  \big],\nonumber\\
\end{eqnarray}
$\text{Si}$ being the sine integral function. The plus sign corresponds to the Dicke-triplet~ $\vert 1,0 \rangle$ state associated with the
$\Sigma^*_u \rightarrow \Sigma_g$ transition, whereas the minus sign corresponds to the transition $\Sigma^*_g \rightarrow \Sigma_g$ which is forbidden for small $\xi$ but becomes allowed for larger $\xi$. 

b) For the $\Pi^* \rightarrow \Sigma (\Delta \Lambda =1)$  transitions one gets the rates



\begin{eqnarray}
     &&\gamma  \pm \frac{1}{t}\int^t_0 \gamma_{|\Delta \Lambda| = 1 }(t') dt' \nonumber\\&&= \frac{\gamma}{t \dot{\xi}}\big[ \xi \pm \frac{3}{4}[\text{Si}( \xi )- \xi^{-1} \cos \xi 
 + \xi^{-2} \sin \xi]  \big],\nonumber\\   
\end{eqnarray}
where the plus or minus sign correspond to a $\Pi^*_u \rightarrow \Sigma_g$ or  $\Pi^*_g \rightarrow \Sigma_g$ transition, respectively.



Thus the Dicke-singlet emission differs appreciably from that of the Dicke-triplet, and so do the $\Delta \Lambda=0$ and $|\Delta \Lambda|=1$ transitions (Fig.~\ref{fig_5}(b)). Importantly, the cooperative contribution $\gamma_{\vert \Delta \Lambda \vert}(t)$ determines the ensemble-averaged ringing rate which results from radiative interference between the emissions of the receding fragments, after then ceases to interact. Hence, such emission can serve as witness of the entanglement of the parent-molecule state long after it has dissociated.

Photoexcitation experiments, such as the one performed by Grangier et al.~\cite{PhysRevLett.54.418}, correspond to Dicke-triplet preparation, whereas a Dicke-singlet, which corresponds to a gerade spin-singlet parent-molecule state, can only be reached by two-photon (Raman) absorption. An alternative is a non-radiative process, e.g. dissociative recombination of $\mathrm{H}_{2}{ }^{+}$with a slow electron to a dissociative ${ }^{1} \Sigma_{g}^{*}$ state.

\subsection{Fluorescence from dissociating alkali dimers}\label{section_5.2}

A qualitatively different type of time-resolved fluorescence from dissociating diatoms has been predicted~\cite{PhysRevA.38.6433}, in cases where the excited atomic fragment is in a \emph{fine-structure} multiplet state, as in alkali atoms. In such cases, the fluorescence exhibits a combination of separation-dependent ringing with quantum beats caused by fine-structure splitting. This effect should arise only under nonadiabatic dynamics that mixes excited adiabatic states which are associated with different fine-structure levels.

The observation of such effects may reveal information on amplitudes of the superposed adiabatic states (\textit{including their relative phase}), and thereby on details of the nonadiabatic mixing producing this superposition. In contrast, \textit{conventional spectroscopy yields only the relative populations of these adiabatic states}.

Let us consider the photodissociation of a homonuclear alkali dimer through the channel ${ }^{1} \Sigma_{u(g)}^{*} \rightarrow{ }^{2} P_{j}^{*}+{ }^{2} S_{1 / 2}$,  conforms to the model of two emitting adiabatic states, each correlated to a single doublet level. The dimer is initialized in the lowest rovibronic state with zero nuclear angular momentum and has the value  $\Omega=0$ for the total angular momentum projection on the molecular internuclear axis, $\Omega=\Lambda+S_{z}$ ($S_{z}$ being the spin projection).

The two adiabatic states with $\Omega=0^{+}$ can then be written at large separations ($R \gtrsim 1 \mathrm{~nm}$) outside the domain of molecular forces, as

\begin{eqnarray}
&& |\mathrm{I}\rangle={}^1\Sigma_{w}^{+*}, \nonumber\\
&& |\mathrm{II}\rangle=\frac{1}{\sqrt{2}}\left[ {}^3\Pi_{w}^{*} (\Lambda = -1, S_z = 1) -  {}^3\Pi_{w}^{*} (\Lambda = +1, S_z = -1)\right].\nonumber\\
\end{eqnarray}
Here $w$ denotes the $u(g)$ symmetry, ${ }^{1} \Sigma_{w}^{+*}$ and ${ }^{3} \Pi_{w}^{*}$ being the  symmetrized superpositions of the product states $|S\rangle_{A(B)}\left|P_{m=\Lambda}\right\rangle_{B(A)}$ multiplied by the appropriate spin functions ( in the $L S$ coupling scheme). 

We next diagonalize the adiabatic Hamiltonian matrix by allowing for their long-range dipole-dipole interactions and the fine-structure splitting~\cite{PhysRevA.38.6433}

\begin{eqnarray}
H_{w}\left(\Omega=0^{+}\right)= \begin{pmatrix} 
 V_{\Sigma} &  \frac{\sqrt{2} }{ 3 }\delta\\
\frac{\sqrt{2}}{ 3 }\delta & V_\Pi - \frac{\delta}{3}
\end{pmatrix},
\end{eqnarray}
where the diagonal terms involve $V_{\Sigma(I)}(R)$,  the dipole-dipole interaction in the state $|\mathrm{I}\rangle(|\mathrm{II}\rangle)$ whereas the off-diagonal term is proportional to $\delta$, the fine-structure splitting.  The resulting  eigenstates $\vert \pm \rangle$ then correlate at separations such that $|V| \gg \delta$, $V=V_{\Sigma}-V_{\Pi}$ (corresponding, e.g. to $R \ll 20 \mathrm{~nm}$ in $\mathrm{Li}_{2}$) to $|-\rangle \rightarrow|\mathrm{I}\rangle\left({ }^{1} \Sigma_{u}^{+*}\right.$ ) or $|\mathrm{II}\rangle\left({ }^{3} \Pi_{g}^{*}\right)$, and to $|+\rangle \rightarrow|\mathrm{I}\rangle\left({ }^{1} \Sigma_{g}^{+*}\right)$ or $|\mathrm{II}\rangle\left({ }^{3} \Pi_{u}^{*}\right)$. In this limit the $\vert - \rangle (\vert + \rangle) $ state is populated for $u ~(g)$ symmetry.   In the opposite limit of large separations $(\delta \gg V)$ 

\begin{eqnarray}
   &&\vert - \rangle \rightarrow \frac{1}{\sqrt{3}}(\vert \mathrm{I} \rangle  - \sqrt{2} \vert\mathrm{II} \rangle ) , \nonumber\\
   &&  \vert + \rangle \rightarrow \frac{1}{\sqrt{3}}( \sqrt{2}\vert \mathrm{I} \rangle  + \sqrt \mathrm{II} \rangle ),
\end{eqnarray}
so that at large separations  $|-\rangle(|+\rangle)$ correlate to  superpositions of the two-atom product states
$\vert^{2} P_{1 / 2(3 / 2)}\rangle_{A(B)}\vert^{2} S_{1 / 2}\rangle_{B(A)}$, that span the basis of their entangled states.




Following non-adiabatic dynamical coupling during dissociation, the emission rate $\dot{P}$, to first order in $\gamma / \delta$, acquires the form~\cite{PhysRevA.38.6433}

\begin{eqnarray}
&&\dot{P}(t) =  \dot{\rho}_{++}+\dot{\rho}_{--} \nonumber\\
&&\simeq  \sum_{ \pm}\gamma\left[\left(1+G_{ \pm}\right) \rho_{ \pm \pm}\left(t_{D}\right) \exp \left[-\gamma \int_{t_{D}}^{t}\left(1+G_{ \pm}\right) d t^{\prime}\right]\right] \nonumber\\ &&+ \gamma A\left(t_{D}\right) G_{c} \cos \left[\int_{t_{D}}^{t} \varepsilon d t^{\prime}+\phi\right] \exp \left[-\gamma \int_{t_{D}}^{t}\left(1+G_{+}+G_{-}\right) d t^{\prime}\right].\nonumber\\
\end{eqnarray}
Here the $\vert \pm \rangle$ populations decay at the rates $\gamma (1 \pm G_{\pm})$ while their cross-term (coherence) decays at the rate $\gamma G_{c}$. This coherence is the entangled two-atom state superposition that has acquired the amplitude $A(t_D)$ and phase $\phi$ during the non-adiabatic dynamical coupling, $t_{D}$ is the delay time required to cross the dynamical-coupling region and the energy splitting of $\vert + \rangle $ and $\vert - \rangle$ is $\varepsilon \simeq \delta$.

The resulting fluorescence rate temporal pattern contains a wealth of information: (1) At low velocities such that $\delta \gg k v$, fast fine-structure beats are superposed on top of the slow ringing envelopes caused by dipole-dipole interactions (Fig.~\ref{fig_6}). The entangled-state (coherence) amplitude $A$ may be deduced from the form of these superbeats, and the populations $\rho_{++}, \rho_{--}$ from the envelope shape. (2)  At intermediate velocities  $\delta \simeq k v$,  the fast superbeat  component oscillates at $2 \delta$ [Fig.~\ref{fig_6}]. (3) The initial rate $\dot{P} / \gamma \simeq 1+w A \cos \phi$, where $w = \pm 1$ corresponds to $u$ or $g$ symmetry, respectively, and can unravel the emerging two-atom entangled state, similarly to that of spinless atoms.

\section{ Teleportation and Entanglement Transfer by Molecular Dissociation}
\label{section_6}
\subsection{Molecular teleportation}
\label{section_6.1}
Following the principle of Vaidman's translational-variable teleportation~\cite{PhysRevA.49.1473}, Opartn\'y and Kurizki proposed to teleport a molecular wavepacket via its collision
with one of the two EPR-entangled atoms emerging from molecular dissociation~\cite{prl_86_3180_2001}. Their proposal was based on the ability of atom pairs formed by diatom dissociation to yield measurable coordinate and momentum correlations that approximate those of the EPR state in Eq.~(\ref{eq: state_1}), i.e. have a large squeezing parameter $s \gg 1$ (Eq.~(\ref{eq_3})). 

\begin{figure*}
 \begin{center}
             \includegraphics[width=0.50\textwidth]{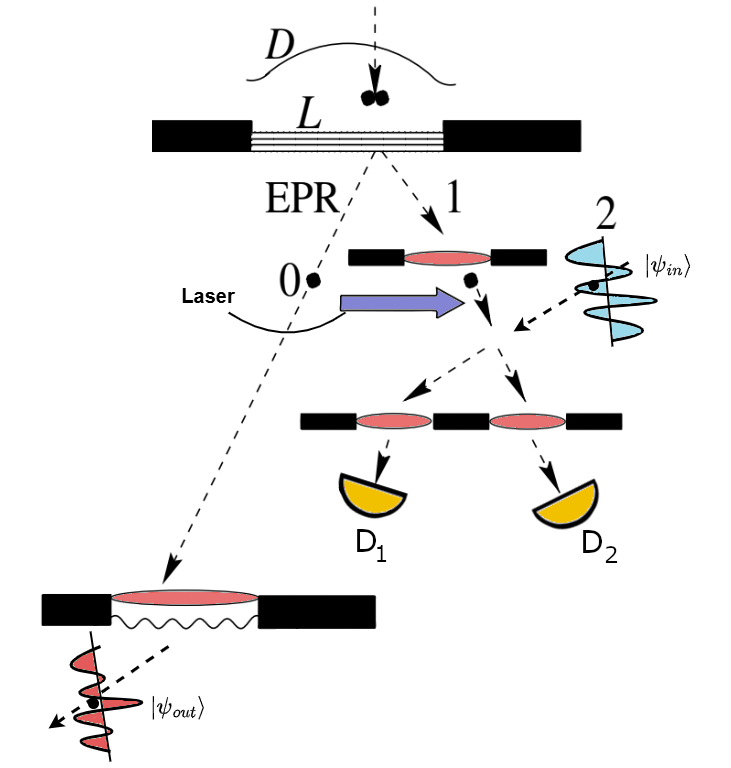}
 \end{center}
 \caption{EPR entanglement and teleportation of atomic or molecular wavepackets. A cold diatom with COM spread $D$ dissociates into translational (EPR) entangled atoms or ions 0 and 1.
Ion 1 is focused and laser-deflected to collide with ion 2 (synchronized events). Their post-collision distance and momentum sum are measured by
detectors 1 and 2 and determine the position and momentum shifts
of atom 0, whose translational state $\vert \psi_{out} \rangle$ then approximately
 reproduces $\vert \psi_{in} \rangle$  of ion 2.}  
 \label{fig_7} 
 \end{figure*}

A concrete proposal for such a teleportation scheme may involve cold ionized molecules that move fast along $z$ to a region where the molecules are dissociated by a laser pulse. 
The dissociation region along the orthogonal x axis is defined by an aperture (Fig.~\ref{fig_7}). The molecular center-of-mass (c.m.) wavepacket along x should approximate the minimum-uncertainty gaussian state of momentum and position prior to dissociation.
Such a state can be prepared by cooling the molecule (of mass $M$) in a trapping potential, 
to a temperature $k_B T \approx \hbar^2/(M D^2)$, where $D$, the c.m. wave packet
size, should satisfy $D \lesssim L$. A size $D \approx 300 \text{nm}$, which
requires 
$T \approx 0.4 \mu \text{K}$ for $\text{Li}^-_2$ is achievable by Raman photoassociation of Li atom pairs in optically-trapped Bose condensates~\cite{rom2004state}. 

In order to teleport an input state $\vert \psi_{in} \rangle$  of a molecular wavepacket 2 to atom 0, wavepacket 2 collides with atom 1 of the 1 and 0 atomic EPR pair, and then the collision partners 1 and 2 are detected. This collision of particles 1 and 2 projects their joint state onto the  EPR-correlated basis states, which are characterized by a separation $x_{rel}$
and momentum sum $p_{c.m.}$. Ideally, the output translational
state $\vert \psi_{out} \rangle$ of atom 0 (which is the other member of the EPR correlated pair) is expected to be nearly identical to the input state $\vert \psi_{in} \rangle$ of the molecular wavepacket 2, if the teleportation is successful.

The teleportation fidelity is determined by the final Wigner distribution $W_{out}(x_0, p_{x0})$  of the output wavepacket 0, which is given by that of the input wavepacket, $W_{in}(x_2, p_{x2})$, convoluted with a smoothing function whose width is determined by the process errors caused by technical limitations or by decoherence,
$\Delta x_{E}$ and $\Delta p_{E}$~\cite{prl_86_3180_2001,fisch2005translational}: 
\begin{subequations}
\begin{eqnarray}
&&W_{out}(\alpha_0) = \int d^2 \alpha W_{in}(\alpha)G_{\sigma}(\alpha_0 - \alpha), \nonumber\\
\end{eqnarray}
where  $\alpha \equiv x + ip$  is a complex variable and $G_{\sigma}(\alpha)$ is a Gaussian of width 
\begin{equation}
    \sigma = e^{-2s_E}, \quad s_E \equiv \hbar/(2 \Delta x_{E} \Delta p_{E}).
\end{equation}
\end{subequations}
When the process errors render the guassian width comparable to the input wavepacket width, teleportation fails, being unable to reproduce the characteristic features of the input state.


\subsection{Molecular entanglement transfer from dissociation fragments to photons}
\label{section_6.2}

\begin{figure*}
 \begin{center}
  \includegraphics[width=0.7\textwidth]{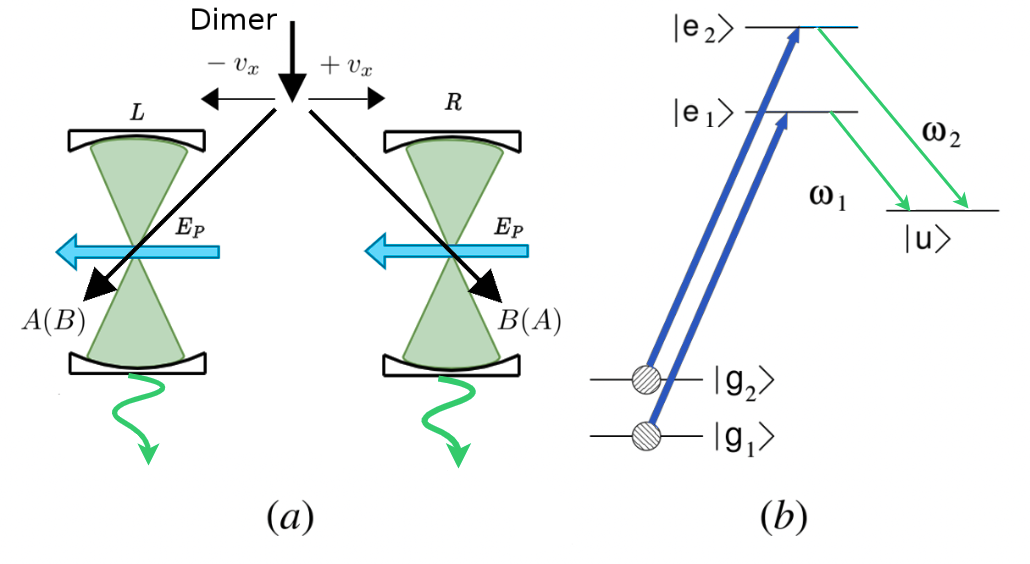}    
 \end{center}
 \caption{(a) Dimer dissociation products $A$ and $B$ traverse the cavities $L$  and $R$, respectively, where they interact with pump fields $E_0$  and emit an entangled (TIE) pair of photons that leak from the cavities. (b) Level scheme of dissociating fragments.}
 \label{fig_8} 
 \end{figure*}

One may transfer the quantum state of two  \textit{molecular-dissociation fragments that are internally- and translationally-entangled} to an entangled state of two photons by the scheme
depicted in  Fig.~\ref{fig_8}~\cite{PhysRevA.67.012318}. Let us assume that  cold diatomic molecules (e.g., $\text{Na}_2$) undergo dissociation via stimulated Raman adiabatic passage (STIRAP)~\cite{RevModPhys.70.1003} ending up in a  state in which the
constituent atoms, $A$ and $B$ 
populate one of the ground-state hyperfine states  $\vert g_{1(2)} \rangle$. In the c.m. frame, 
these atoms recede along the x axis with velocities
$\pm v_x$, that are dependent on their hyperfine state. 
Two high-$Q$ cavities, $L$ and $R$,  are positioned (Fig.~\ref{fig_8}) so as to allow only a pair of  atoms in a state $\vert g_1 \rangle_A \vert g_2 \rangle_B$ or 
$\vert g_2 \rangle_A \vert g_1 \rangle_B$ to traverse both cavities, while discarding other events. The cavity
fields modes overlap with laser pump fields $E_{p_{1,2}}$ that have
the frequencies $\omega_{p_{1,2}}$ and wavevectors $k_{p_{1,2}}$ along $x$. In each
cavity the two modes are split in frequency by $(\omega_{p_1}-\omega_{p_2})+(\omega_{g_1}-\omega_{g_2})$, while the L and R cavity-mode frequencies differ by  
$2k_{p_{1,2}} v_x$, the difference between the Doppler shifts of the
photons generated in the two cavities. This condition
requires choosing the appropriate
intermediate excited states $\vert e_1 \rangle$ and $\vert e_2 \rangle$ in Fig.~\ref{fig_8} (such as $4\text{P}_1/2$ and $4\text{P}_3/2$ in sodium).

STIRAP transfer~\cite{RevModPhys.70.1003} to the final state $\vert u \rangle$ ($4\text{S}_1/2$ in Na) emits a photon in each cavity at one of the admissible frequencies, assuming strong coupling of the atoms with the cavity fields. When these photons leak out the cavity, the atom-pair entangled state formed by dissociation, is mapped to an entangled photon-pair state:

\begin{eqnarray}\label{eq28}
 (\vert g_1, - v_x \rangle_A \vert g_2, v_x \rangle_B \pm \vert g_2, - v_x \rangle_A \vert g_1, v_x \rangle_B  \vert 0 \rangle_L  \vert 0 \rangle_R) \nonumber\\
 \rightarrow \vert u, -v_x\rangle_A \vert u, v_x\rangle_B (\vert \omega_1 \rangle_L \vert \omega_2 \rangle_R \pm \vert \omega_2 \rangle_L \vert \omega_1 \rangle_R). \nonumber\\    
\end{eqnarray}
Here, the symmetry (parity) of this entangled photonic state and the frequencies $\omega^{(L,R)}$ encode information on the initial, \textit{translationally-internally entangled} (TIE) molecular state.

The high-fidelity entanglement transfer from the dissociated fragments to the emitted photon pair can be followed by \emph{the reversal of the process in a pair of cavities at a distant location}, thus recreating the initial dissociation molecular state at this node (the left hand-side of Eq.~\eqref{eq28}). The described protocol has the potential of moving the quantum state of a composite molecular object between distant nodes of a quantum network~\cite{PhysRevLett.78.3221,kimble2008quantum,kurizki2015quantum}. This protocol can transfer more quantum information than the teleportation of discrete, continuous or TIE variables~\cite{PhysRevLett.80.869,furusawa1998unconditional,PhysRevLett.78.3221,parkins1999quantum} as it teleports \textit{TIE states}~\cite{kolavr2007path}.


  \section{Quantum Thermodynamics of Entangled Dimers}
 \label{section_7}

    \begin{figure*}
 \begin{center}
  \includegraphics[width=0.6\textwidth]{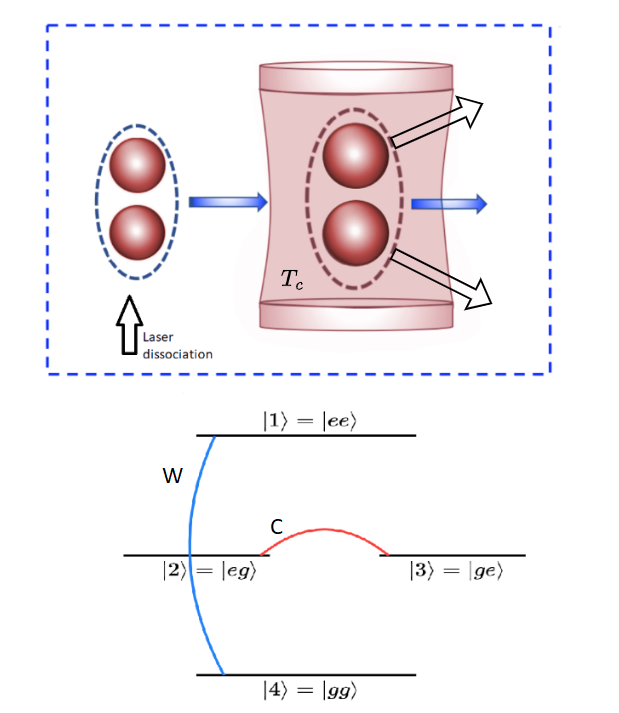}   \end{center}
 \caption{A cavity resonantly pumped by atom pairs formed via dimer dissociation by a laser beam at the entrance to the cavity. The atoms pairs determine the temperature $T_c$ that of the cavity field through the population inversion $w \coloneq 2(\rho_{11} - \rho_{44})$ and the entangled-state coherence $C \coloneq 2 \text{~Re~} \rho_{23}.$}
 \label{fig_9} 
 \end{figure*}

     \begin{figure*}
 \begin{center}
  \includegraphics[width=0.7\textwidth]{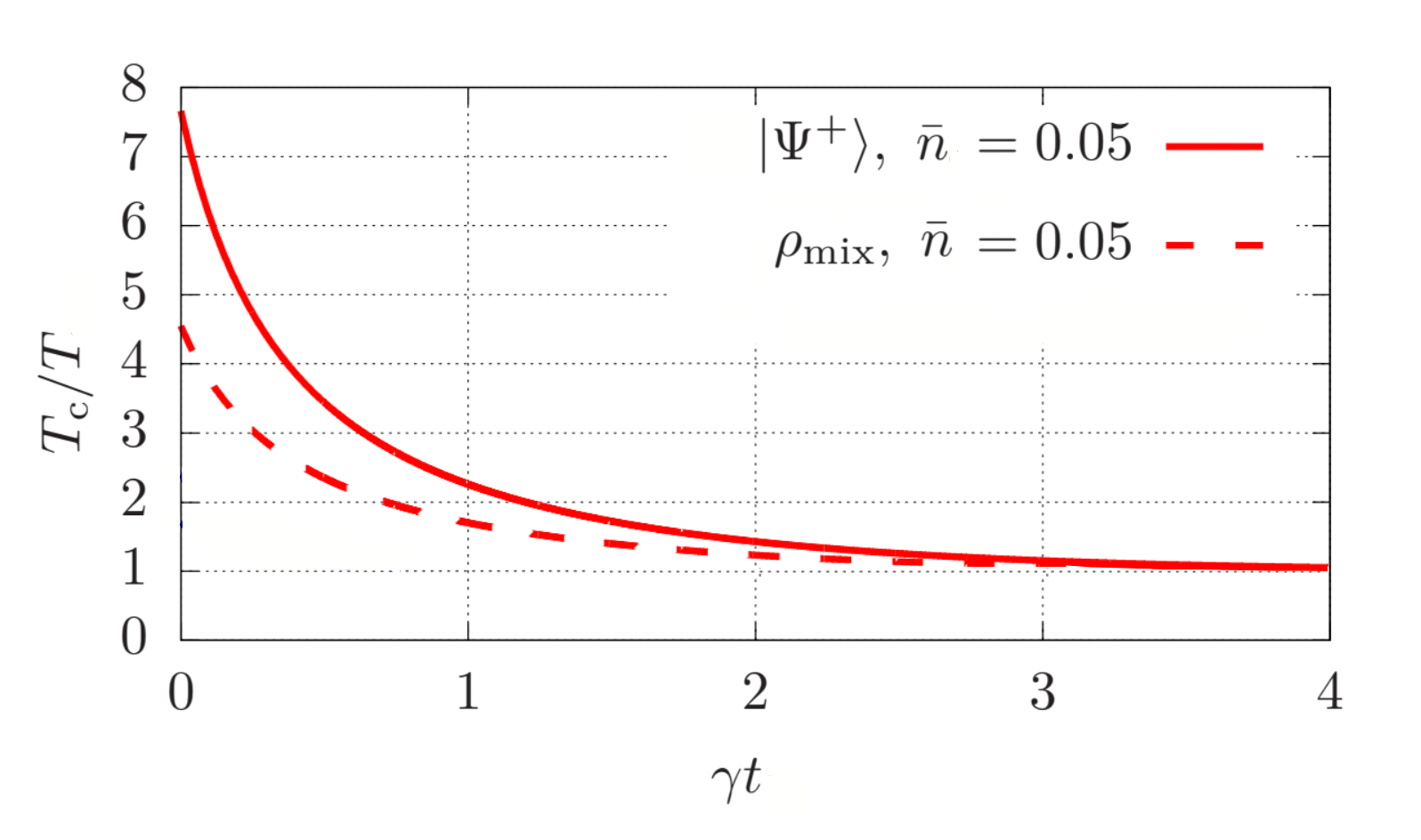}   \end{center}
 \caption{Cavity temperature enhancement by dimer-state entanglement prior to dissociation as a function of the atom-pair transfer time through the cavity, (in units of their decay time $1/\gamma$). The cavity with $\bar{n} = 0.05$ photons.  The excited atoms decay at a rate $\gamma$. The initial state of the dimer is either the bright Dicke state $\vert \Psi^+\rangle$  or $\rho_{\text{mix}}$. The effective cavity temperature $T_c$, scaled by the environment temperature $T$, is strongly enhanced for $\gamma t \ll 1$. }
 \label{fig_10} 
 \end{figure*}

A landmark paper by Scully et al. showed the possibility of using \textit{atomic coherence as a  ``temperature knob"} of a thermal bath~\cite{scully2003extracting} 
thereby opening the way to the new area of quantum thermodynamics~\cite{binder2018thermodynamics,scully2011quantum,niedenzu2018quantum,cangemi2024quantum,tc_2}. In addition to its 
applications in quantum heat machines~\cite{tc_2}, the ability of quantum coherence to engineer thermal baths offers fundamental insights
into the role of quantumness in energy exchange processes~\cite{PhysRevLett.54.418}.

The question we posed was: \textit{Is there a difference between the quantum thermodynamics of single-atom coherence and that of two-atom entanglement?} To this end, we have examined the temperature control of a leaky cavity field mode
that interacts with a quantum-entangled ensemble of dimers. We have shown~\cite{dag2019temperature,daug2016multiatom} that their \textit{two-atom entanglement can strongly enhance the heat exchange of the dimers with the heat bath} compared to analogous effects caused by single-atom coherence~\cite{scully2003extracting,scully2011quantum}.

Dimers injected into a cavity (Fig.~\ref{fig_9}) are described by a state spanned in the basis 

$$\vert 1\rangle = \vert ee\rangle, \vert 2\rangle = \vert eg\rangle , \vert 3\rangle = \vert ge\rangle, \vert 4\rangle = \vert gg\rangle $$

$\vert g \rangle$ and $\vert e \rangle$ denote respectively the lower 
and upper states of the atoms obtained by dimer dissociation. The dimer state is quantum-entangled, provided~\cite{PhysRevLett.77.1413} $\vert \rho_{23}\vert^2 > \rho_{11} \rho_{44}$. We showed~\cite{dag2019temperature,daug2016multiatom} that the \textit{entanglement/coherence element} 
 $C\equiv \rho_{23}$, \textit{can modify the effective temperature} of the cavity-field mode. 

According to micromaser theory~\cite{PhysRevA.34.3077,scully1997quantum,meystre2007elements}, the master equation which governs the cavity-field state that resonantly  interacts with the dimers has the form~\cite{daug2016multiatom}

\begin{eqnarray}
  \dot{\rho}_c = \Big[ \frac{g_{eff}(1+c-w)}{2} + \eta (\bar{n} + 1)  \Big] \mathbf{L}_d \rho_c \nonumber \\
  +  \Big[ \frac{g_{}(1+c+w)}{2} + \eta \bar{n}  \Big] \mathbf{L}_e \rho_c.
\end{eqnarray}
Here $g_{eff}$ is the effective dimer-cavity coupling rate,  $\mathbf{L}_d$ and $\mathbf{L}_e$  denote the Lindblad superoperators~\cite{breuer2002theory} for decay and excitation of the cavity field, respectively, $\eta$  is the cavity leakage rate, and $\bar{n}$ is the population of the thermal photons at temperature $T$ in the absence of dimers. The dimer affects the cavity field via the excitation and the deexcitation coefficients 

\begin{eqnarray}
    r_{\pm} \coloneq 1 + C \pm \frac{w}{2}.
\end{eqnarray}
 where $w \coloneq 2(\rho_{11} - \rho_{44})$ is the population inversion of the doubly-excited state, ranging from $w= -2$ to $+2$ and the two-atom coherence/entanglement parameter is $C \coloneq 2 \text{~Re~} \rho_{23}$.

Remarkably, $T_c$ turns out to be the \textit{effective temperature} for the cavity field that is controllable by the two-atom entanglement coherence $C$. Namely, the dimers act as \emph{a heat bath} at the cavity-field temperature $T_c$. 

The condition $T_c > T$ signifies heating compared to the environment temperature $T$. This heating regime  holds for~\cite{dag2019temperature,daug2016multiatom}

\begin{eqnarray}
  C > -1 - (\bar{n} + \frac{1}{2})w.  
\end{eqnarray}
This inequality shows that the heating caused by entanglement is maximal when the environment temperature is very low.  The single-excitation populations $\rho_{22}$ and $\rho_{33}$ nevertheless determine the maximally allowed coherence $C$ through the condition $\vert C \vert \le 2 \sqrt{\rho_{22} \rho_{33}}$ which can be recast as

\begin{eqnarray}
  \vert C \vert \le 1 - \frac{\vert w \vert}{2}.  
\end{eqnarray}

Hence, the two-atom coherence/entanglement $C$ determines, for a given dimer excitation energy, which is proportional to $\vert w \vert$,  what would be the thermal steady state of the cavity field. Thus, the two singly-excited Bell states

\begin{subequations}
 \begin{equation}
    \vert \Psi^{\pm}\rangle \coloneq  \frac{\vert ge \rangle \pm \vert e g \rangle}{\sqrt{2}}
 \end{equation}  
have the \emph{same energy} as their phase-averaged counterpart
  \begin{equation}
    \rho_{\text{mix}} \coloneq \frac{1}{2} \vert ge \rangle \langle ge \vert  + \frac{1}{2} \vert eg \rangle \langle eg \vert
 \end{equation}  
\end{subequations}
yet each of them gives rise to a very different temperature

\begin{eqnarray}
   T_+ > T_{\text{mix}} > T_- \equiv T 
\end{eqnarray}

where $T_{\pm}$ pertains to $\vert \Psi^{\pm}\rangle$ ($C = \pm 1$) and $T_{\text{mix}}$  to $\rho_{\text{mix}}$  ($C = 0$). 

The Bell state $\vert \Psi^+ \rangle$ yields higher temperature $T_c$ (than other states for which $w=0$). By contrast, $\vert \Psi^{-}\rangle$ does not change the cavity temperature relative to $T$. This stems from the properties of the superradiant Bell state $\vert \Psi^+ \rangle$ that yields enhanced energy transfer from the atoms to the cavity via constructive interference as opposed to the subradiant $\vert \Psi^{-} \rangle$ state, whose destructive interference prevents any such transfer. As shown in Fig.~\ref{fig_10}, superradiant-state entanglement can strongly enhance the effective cavity temperature under realistic conditions. As discussed in Sec.~\ref{section_5.1} $\vert \Psi^-\rangle$ or $\vert \Psi^+ \rangle$ are obtained by two-photon or single-photon dimer-state dissociation, respectively (Fig.~\ref{fig_5}a). 

To conclude, our studies~\cite{dag2019temperature,daug2016multiatom} have shown that entangled two-atom states obtained by dimer dissociation, can be a heating resource unless they are gravely corrupted by dissipation effects due to cavity leakage and dimer decoherence. In fact, such entangled atom pairs are able to provide cavity temperature control over
a much broader temperature range than single-atom coherent quantum baths. Owing to this property, entangled diatoms can serve as advantageous fuel in photonic heat engines and endow them high efficiency.


 \section{Discussion}
 \label{section_8}

We have presented our perspective of molecular processes that exhibit entanglement and can thus serve as resources of quantum information \rr{and its processing for two general purposes: i) diagnostics of possibly unknown molecular quantum states and their dynamics via the relation between their time-resolved cooperative fluorescence and the molecular-state symmetry and angular momentum; ii) teleportation of molecular states that pertain to discrete,  continuous or translationally-internally entangled (TIE)  degrees of freedom of the molecule. The protocols described here are not merely conceptual; they are based on a realistic, experimentally verifiable description of the processes involved that are further detailed in the corresponding references.}

\rr{As in any quantum information processing scheme, measurements are required for the verification and utilization of the surveyed protocols:}

 a) In molecular collisions, the scattering resonance width and the initial wavepackets of the collision partners have been shown to determine the degree of their translational entanglement~\cite{PhysRevLett.94.160503} (Fig.~\ref{fig_2}, Sec.~\ref{section_3}). \rr{Scattering cross-section measurements~\cite{baer2012molecular,child2014molecular} can provide the required information for quantifying the collisional entanglement.} \rrr{These measurements are destructive: After the collection of this information, the entangled state no longer exists.}

 b) In molecular Raman dissociation \rr{(Fig.~\ref{fig_3})}, one may control the EPR entanglement by choosing appropriate laser characteristics~\cite{fisch2005translational} (Sec.~\ref{section_4.1}). Results of two-atom global (joint) coordinate- and momentum measurements of the entangled atoms, which are an \textit{essential part of the teleportation protocol}~\cite{prl_86_3180_2001}, can be inferred from the positions of the atomic detectors and their angular separation, respectively (Fig.~\ref{fig_7}). \rrr{These measurements cannot be deemed destructive, as they serve the purpose of enabling teleportation.}

 c) Double-well tweezers may host the dissociation partners and facilitate the measurement of their EPR correlations~\cite{fisch2005translational} (Sec.~\ref{section_4.2}). However, the shallowness of optically-induced trap potentials (typically $\approx$ 1mK) drastically
limits the momentum uncertainty of atoms that can be stored to $\Delta p_{rel} \ll \sqrt{2mV_{\text{opt}} }$ where $V_{\text{opt}}$ is the trap potential depth. \bbb{As a substitute for tweezers, one may coherently and continuously monitor the receding wavepackets by off-resonant photon scattering~\cite{PhysRevA.87.052141}, \rrr{that constitute non-demolition (non-destructive) measurements}~\cite{scully1997quantum,meystre2007elements}. }


 d) The fluorescence emitted by dissociating molecular fragments may act as an entanglement witness that can disclose the parity, angular momentum, and dissociation dynamics of the parent molecule~\cite{PhysRevA.32.2560,PhysRevA.36.90,ben1987stereospecificity,PhysRevA.38.6433} (Sec.~\ref{section_5}). The measurements that provide this information are the same as in the experiment of Grangier, Aspect, and Vigu\'e~\cite{PhysRevLett.54.418}: the fluorescence rate should be recorded with typically nsec resolution.  Sub-nsec resolution is needed to measure fluorescence from fragments that form TIE states, as in dissociating alkali diatoms (Fig.~\ref{fig_6}, Sec.~\ref{section_5.2}). \rrr{These measurements are meant to collect information on the state, rather than preserve it. Hence, they may be considered destructive. }

 e) A variant of such entanglement witness processes is the photoassociation of colliding partners in a cavity: under strong coupling to the cavity field, single-photon absorption may bind the colliding partners in a giant loosely bound diatom with distinct symmetry and angular momentum~\cite{PhysRevLett.83.714}. Such processes \rrr{are conditioned on the outcomes of measurements of photon transmission} on time scales within the cavity transit-times of the atoms, which are typically sub-nsec~\cite{scully1997quantum,meystre2007elements,PhysRevA.87.052141}. 

f) The above processes and measurements are all feasible using existing experimental techniques. They are prerequisites for the teleportation of a molecular wavepacket (Sec.~\ref{section_6.1}) by a scheme~\cite{prl_86_3180_2001} that relies (Fig.~\ref{fig_7}) on measurements realizable by ion optics, laser-induced molecular dissociation, and laser control of collisions with adequate precision.

g) An alternative to teleportation is the scheme (Fig.~\ref{fig_8}, Sec.~\ref{section_6.2})  of entanglement transfer via STIRAP from molecular  TIE states in a cavity to entangled photons that can subsequently recreate those states in another distant cavity~\cite{PhysRevA.67.012318}. The advantage of this scheme, which may be incorporated in protocols of quantum teleportation, cryptography and communications, is that it relies on the STIRAP scheme, which is well within the experimental state-of-the-art~\cite{RevModPhys.70.1003}. Nevertheless, it still faces non-trivial (albeit manageable) experimental hurdles. \rrr{This scheme does not alter the entangled state, and is thus not destructive. }

h) The protocols presented here for molecules apply to trapped cold-atom gases as well, but different experimental and measurement techniques are required there~\cite{PhysRevLett.106.120404,gross2011atomic,PhysRevLett.90.250404,PhysRevLett.127.140402}. One such protocol has already been experimentally implemented~\cite{gross2011atomic}.

To conclude, molecular processes have been shown to store quantum information that may become a potentially useful resource for quantum information protocols. They may raise intriguing questions regarding the role of quantum entanglement in a variety of scenarios involving molecules which are relevant for molecular coherent and entanglement-assisted control~\cite{gong_jcp_2003,gong_jcp_2010,shapiro_rsc_2011,shapiro_prl_2011,shen2019coherent,PhysRevA.78.012334,kim2011ultrafast,lopez2022coherent,ranitovic2014attosecond,lin2020quantum,PhysRevA.76.013409,shobeiry2024emission,PhysRevLett.131.013201,gong2003entanglement,ranitovic2014attosecond,schnabel2025discovery,yurovsky2006formation}. These open questions provide an outlook to further experimental and theoretical work that would span topics from atomic, molecular, and optical physics as well as quantum information science:

1) Can we extend the signatures of dissociative or collisional entangled states that reflect  (Sec.~\ref{section_2}-\ref{section_5}) the symmetry, angular momentum, and energy surface of the parent molecule to polyatomic molecules, rather than the dimers considered here? Such an extension would allow the development of broadly applicable time- and energy-resolved diagnostic tools of molecular states and processes~\cite{mukamel1997electronic,mukamel2000multidimensional}.

2)  Can entanglement considerations be used to select specific molecular reaction channels via feedback that is driven by the outcomes of the measurements discussed above (in a)-g))? This approach would be an alternative to the existing molecular coherent-control schemes~\cite{gong_jcp_2003,gong_jcp_2010,shapiro_rsc_2011,shapiro_prl_2011}.

3) Can we extend the teleportation and STIRAP-based schemes in Sec.~\ref{section_6}  to more complex scenarios involving molecules with many degrees of freedom? This would be a major step towards teleporting complex molecules and using them as building blocks of a remotely assembled large object,  a challenging goal that may nevertheless become reachable in the future.

4) Can we extend the scheme described in Sec.~\ref{section_7}  to the design of heat engines that would extract work from entanglement or store it (in the quantum battery configuration~\cite{alicki2013entanglement,binder2015quantacell,campaioli2017enhancing,garcia2020fluctuations}) with high efficiency? What would be the practical limits on the efficiency that is boosted by the entangled state of the molecular complex?


\section*{Acknowledgments} 
GK acknowledges the support of NSF-BSF and DFG (FOR 2724). PC acknowledges support from the International Postdoctoral Fellowship from the Ben May Center for Theory and Computation.


%

\end{document}